\documentclass[10pt]{article}
\usepackage{geometry}
\usepackage{fullpage}
\usepackage{graphicx}
\usepackage{amsmath}
\usepackage{amssymb}
\usepackage{multirow,tabularx}
\usepackage{subfig}
\usepackage{hyperref}

\title{ Protein Conformational States: A First Principles Bayesian Method}
\author{
David M. Rogers\\
Chemistry and Materials Group\\
 National Center for Computational Sciences\\
Oak Ridge National Laboratory}

\begin{document}
\maketitle
\begin{abstract}
  Automated identification of protein conformational states
from simulation of an ensemble of structures is a hard problem
because it requires teaching a computer to recognize shapes.
We adapt the na\"{\i}ve Bayes classifier from the machine learning community
for use on atom-to-atom pairwise contacts.
The result is an unsupervised learning algorithm that
samples a `distribution' over potential classification schemes.
We apply the classifier to a series of test structures
and one real protein, showing that it identifies the
conformational transition with $>$ 95\% accuracy in most cases.
A nontrivial feature of our adaptation is a new connection
to information entropy that allows us to vary the level of
structural detail without spoiling the categorization.
This is confirmed by comparing results as the number of
atoms and time-samples are varied over 1.5 orders of magnitude.
Further, the method's derivation from Bayesian analysis on
the set of inter-atomic contacts makes it easy to
understand and extend to more complex cases.
\end{abstract}

\vspace{2em}
{\small \centering

This manuscript has been authored by UT-Battelle, LLC, under contract DE-AC05-00OR22725 with the US Department of Energy (DOE). The US government retains and the publisher, by accepting the article for publication, acknowledges that the US government retains a nonexclusive, paid-up, irrevocable, worldwide license to publish or reproduce the published form of this manuscript, or allow others to do so, for US government purposes. DOE will provide public access to these results of federally sponsored research in accordance with the DOE Public Access Plan (http://energy.gov/downloads/doe-public-access-plan).}

\clearpage\twocolumn\sloppy

\section{ Introduction}

  The conventional description of protein dynamics asserts that proteins
posses intrinsic conformational states.\cite{jguo16}
An enzyme may cycle between catalytic
and open states.\cite{cmull96}
An ion channel may open and close its central pore.\cite{camar12}
A chaperone protein assists transformation of large, hydrophobic proteins
from from initial, linear, to folded shapes.\cite{ykim13}
X-ray and cryo electron-microscopy
reveals conformations with small motions on the 1-5 \AA{}ngstrom level for those
proteins that crystallize.\cite{jvant20}
Neutron scattering and nuclear magnetic resonance
structures of room temperature proteins show greater shape variability, but are
usually able to classify structures into a few `canonical' structures.

  Advances in molecular modeling have made it possible to simulate
the protein folding process, generating very large numbers of samples,
free energy landscapes, and information on kinetics.  
Nevertheless, computational identification of distinct conformational
states from molecular simulations has remained an active area of methodological research.

  One of the principle methods developed for visualizing domain motions
in proteins is DynDom\cite{rlee03,cgird15}.
DynDom works from two input structures
and determines relative domain rotations.
This allows predicting transition motions from experimentally observed
conformers, but not conformations from observed motions.
On the other hand, many structure-to-structure similarity
classification methods have emerged for this problem.\cite{hstam10,arama11,aferg11}
A recent review of such dimensionality reduction methods for
protein conformational spaces noted that nonlinear
methods are generally better than Cartesian or linear ones,
but that the complexity of assumptions behind
those models makes them difficult
to work with and adapt.\cite{dmoji13,bspiw20}

  This work presents a complete inference method derived from one single
statistical hypothesis: that conformational states are defined by
sets of contacting residues.  Specifically, the conformational state, $k$,
uniquely determines which pairs of residues $u$,$v$,
will be touching.  Like a weighted coin flip, the contact
probability is $\mu_{k,(u:v)}$
-- independently from all the other contacting pairs.
Each conformational state is thus characterized by a
vector, $\mu_k$, encoding the set of contacting
pairs in state $k$.

  The statistical model derived from this problem statement
is termed a Bernoulli mixture model for binary feature classification.\cite{cli16}
The problem setup is similar to the Naive Bayes method.\cite{valab06}
However, because the categories are not known in advance, this is an
unsupervised learning and classification problem.

  Bernoulli mixture models have been applied extensively
in the field of text subject analysis,\cite{jnovo04}
optical character recognition,\cite{valab06}
and image feature classification.\cite{cli16}
Essentially all of these applications have been
successful at building extremely accurate classification
models.  The latter work also presents a thorough
summary of sampling methods.

  However, there remain difficulties sampling the distribution
over categories, $\mu$, especially when the number of categories
and reference classifications are not known in advance.
The well-known expectation-maximization algorithm (EM)\cite{dkaji10}
is available in principle, but is not a replacement for sampling.
Theoretical work on the EM method\cite{kyama13}
shows that redundant categories will result in many circumstances.
In this work, we have introduced a prior that eliminates
redundant categorizations.

  This work presents a novel sampling method for category
inference, along with results demonstrating that it creates
structurally meaningful categories with $>$ 90\% accuracy.
Although the potential application space is vast,
this work focuses on proving method robustness
using well-defined synthetic test problems.
Each follows a time sequence mimicking domain motions
in proteins -- so that the classification accuracy can be judged by
correctly assigning categories in time-order.

\section{ Theory}

  A na\"{\i}ve Bayes model (for bit-strings) assumes that structural input samples,
$i=1,\ldots,N$ are generated by first selecting a conformational state,
$z_i \in \{1,\ldots,K\}$,
with probability $\pi_{z_i}$, and then independently
deciding whether each point-to-point contact ($x_{ij}, j=1,\ldots M$),
is made with probability $\mu_{z_i j}$
If contact $j$ is made in sample number $i$, then $x_{ij} = 1$.
Otherwise $x_{ij} = 0$.
The model parameters are $\theta = (K,\pi,\mu)$.

  It leads to a sampling distribution,
\begin{align}
P(x z | \theta) &= \prod_i \pi_{z_i} \prod_{j=1}^F \mu_{z_i j}^{x_{ij}} (1-\mu_{z_i j})^{1-x_{ij}} \\
 &= \prod_{k=1}^K \pi_k^{N_k} \prod_{j=1}^M
\mu_{kj}^{N_{kj}} (1-\mu_{kj})^{N_k-N_{kj}}
.\label{e:like}
\end{align}
The second line above notes that, once the categorization, $z$, is known,
the sampling distribution is easy to express in terms of feature counts
in $\{i : z_i = k\}$ -- the set of samples assigned to category $k$,
\begin{align}
N_k &= |\{i : z_i = k\}| , &
N_{k j} &= \sum_{\{i : z_i = k\}} x_{ij}
.
\end{align}
The first is the number of samples in set $k$, and the second
is the number of times each contact is seen in that set.

\begin{figure}
{\centering
\includegraphics[width=0.45\textwidth]{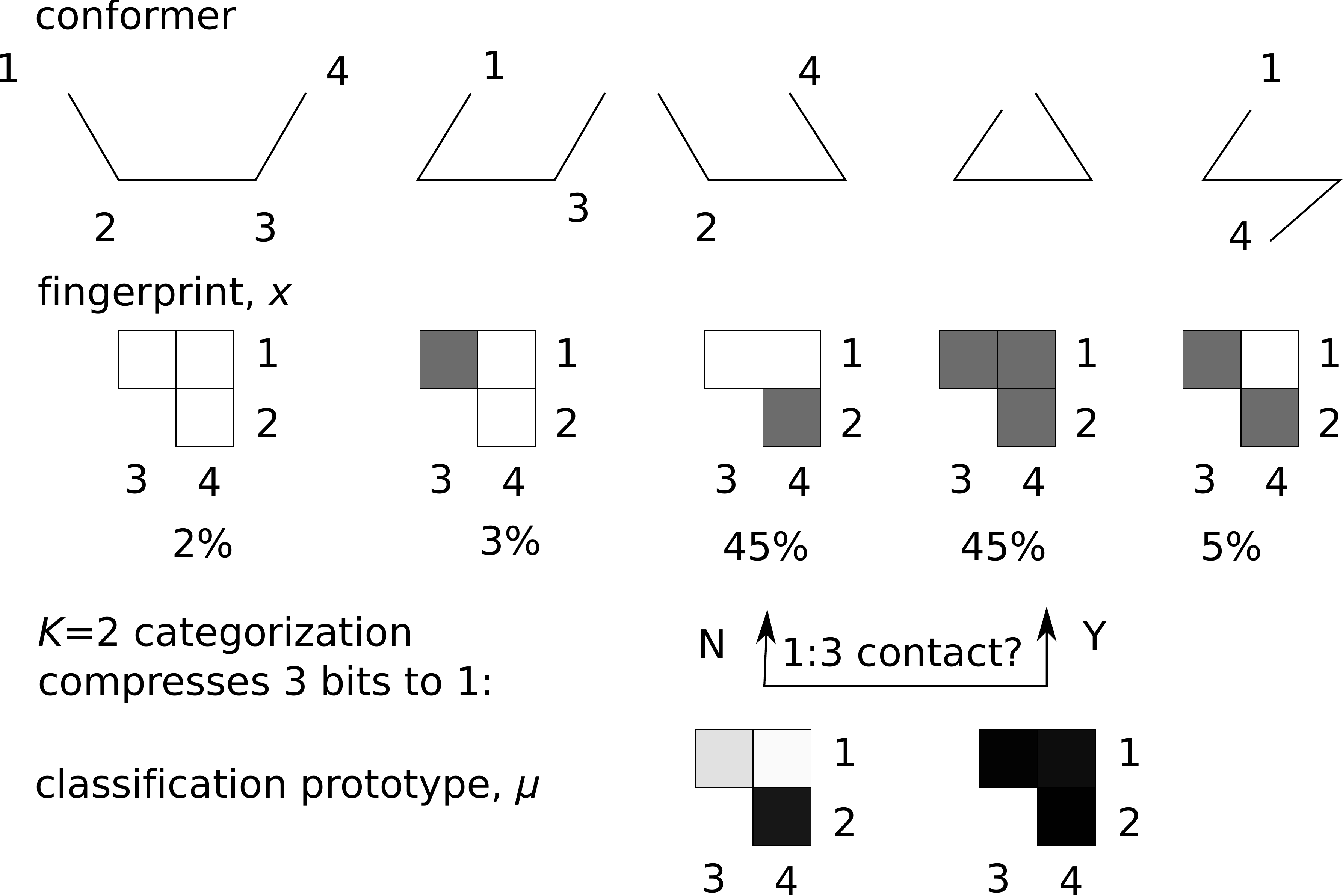}
\caption{Classification applied to a 4-site structure.
Pairs of contacting points are mapped to binary fingerprints, $x$.
The bottom images show the 2-category classification
result for the inputs shown.
Our category split MC move using (1:3) [or 1:4] succeeds here
because those contacts are correlated with others.
The small positive probability for 1:3 contacts
hints that the left structures should be lumped with conformation 1.
}\label{f:scheme} }
\end{figure}

  According to Bayes' theorem,\cite{ejayn03} we can turn this around
to predict two important things -- the probability that sample $i$
belongs to category $k$, (read $z$ given $x$ and $\theta$),
\begin{equation}
P(z_i = k|x_i \theta) \propto \prod_{j=1}^M \mu_{kj}^{x_{ij}} (1-\mu_{kj})^{1-x_{ij}}
,
\end{equation}
and also the probability distribution over all possible parameters,
\begin{equation}
P(\theta z | x I) = C(x) P(x z | \theta) P(\theta | I)
,
\end{equation}
where $C(x)$ is an $x$-dependent normalization constant.
Sampling this distribution provides everything --
the categorizations, $z$, the conformational states,
$\pi, \mu$, and even a predicted number of categories, $K$.

  We choose a prior probability,
\begin{align}
P(\theta|I) &\propto P(\theta | U)
		 \prod_{k < l} \left( 1 - B(\mu_k,\mu_l) \right)
	, \\
P(\theta | U) &= \frac{\Gamma(K \alpha)}{\Gamma(\alpha)^K}
		 \prod_{k=1}^K \pi_k^{\alpha - 1}
		 \label{e:prior}
\end{align}
that enforces uniqueness of the categories.
Here $B(\mu_k, \mu_l) = B_{kl}$ is the Bhattacharyya
similarity between distributions $p$ and $q$,
\begin{equation}
B(p, q) = \prod_{j=1}^M \sqrt{p_j q_j} + \sqrt{(1-p_j)(1-q_j)}
.
\end{equation}
If two categories share the same distribution,
then $p = q$, and $B(p, q) = 1$.  This forces our estimator
to return zero likelihood that $\mu_k = \mu_l$ for any $k\ne l$.

  The second part of Eq.~\ref{e:prior}, $P(\theta|U)$, is a
conventional prior used for Bernoulli mixture inference in the literature.
We use $\alpha = M+1$ throughout this work.
The Supplementary information contains a detailed
justification for this choice.
Essentially, it forces the likelihood for dividing a
category into two parts to be asymptotically
insensitive to the number of features, $M$.
A proof of this fact, as well as a useful
connection to the information entropy of compression
is present in the supplementary information.

\subsection{ Sampling Method}

  Our sampling method is traditional Markov Chain Monte Carlo
using four types of moves: a recategorization move, where
categories, $z$, are assigned according to $P(z|\theta x I)$,
a reclassification move, where $\theta$ is sampled from
$P(\theta | z x U)$ and accepted with probability
$\prod_{k < l} (1-B_{kl})$, and one split and one
join rule.  The function, $P(\theta | z x U)$, referred to here
is just $P(\theta |z x I)$ without the Bhattacharyya distance terms,
and with a different constant prefactor.

    Generating split or join trial moves was done by randomly
choosing either one category to split or choosing
two categories to join.  For splits, member $i$ of category $k$ is
moved to set $L$ with probability
$\eta^{x_{ij}} (1-\eta)^{1-x_{ij}}$.  We used $\eta = 0.9$,
but any $\eta \in (0.5, 1.0)$ should work in principle.
If all elements end up in $L$ or $R$, the partitioning is re-done.
To concentrate splitting on productive cases, we did not
attempt to split categories with $N_{kj} = 0$ or $N_{kj} = N_k$.
Immediately after splitting or joining categories, a reclassification
move (re-assigning $\theta$) was performed.
Category split moves were accepted using the Metropolis criterion,
which is the smaller of
$P_\text{gen,join} P_\text{split} / P_\text{gen,split} P_\text{joined}$ or one.
Explicit formulas for the move generation probabilities ($P_\text{gen,split}$, etc.) are
provided in the supplementary information.

  Fig.~\ref{f:scheme} provides a graphical summary of
this inference scheme.  Each conformational sample is mapped to a bit-string,
which is used as the basis for inferring $\mu$.  Inference
proceeds by sampling potential parameters until a good explanation
for the data is found.  Trial moves that re-categorize and update
$\mu$ look horizontally to find better category prototypes.  Trial moves
that split categories based on presence or absence of
some features allow us to traverse category space vertically.

\section{ Test Systems}\label{s:test}

  The ability of our sampling procedure to predict categories was tested on three
geometrical systems: `chomp' (a closing angle), `heli' (a rotating line), and `glob'
(three rotating spheres).
Each system was generated as a time-series of
1000 frames for approximately $P$ total particles
in 2 or 3-dimensional space.
After generation, Gaussian random noise of width
$\sigma = 0.1$ was added to every degree of freedom.

\begin{figure}
\includegraphics[width=0.45\textwidth]{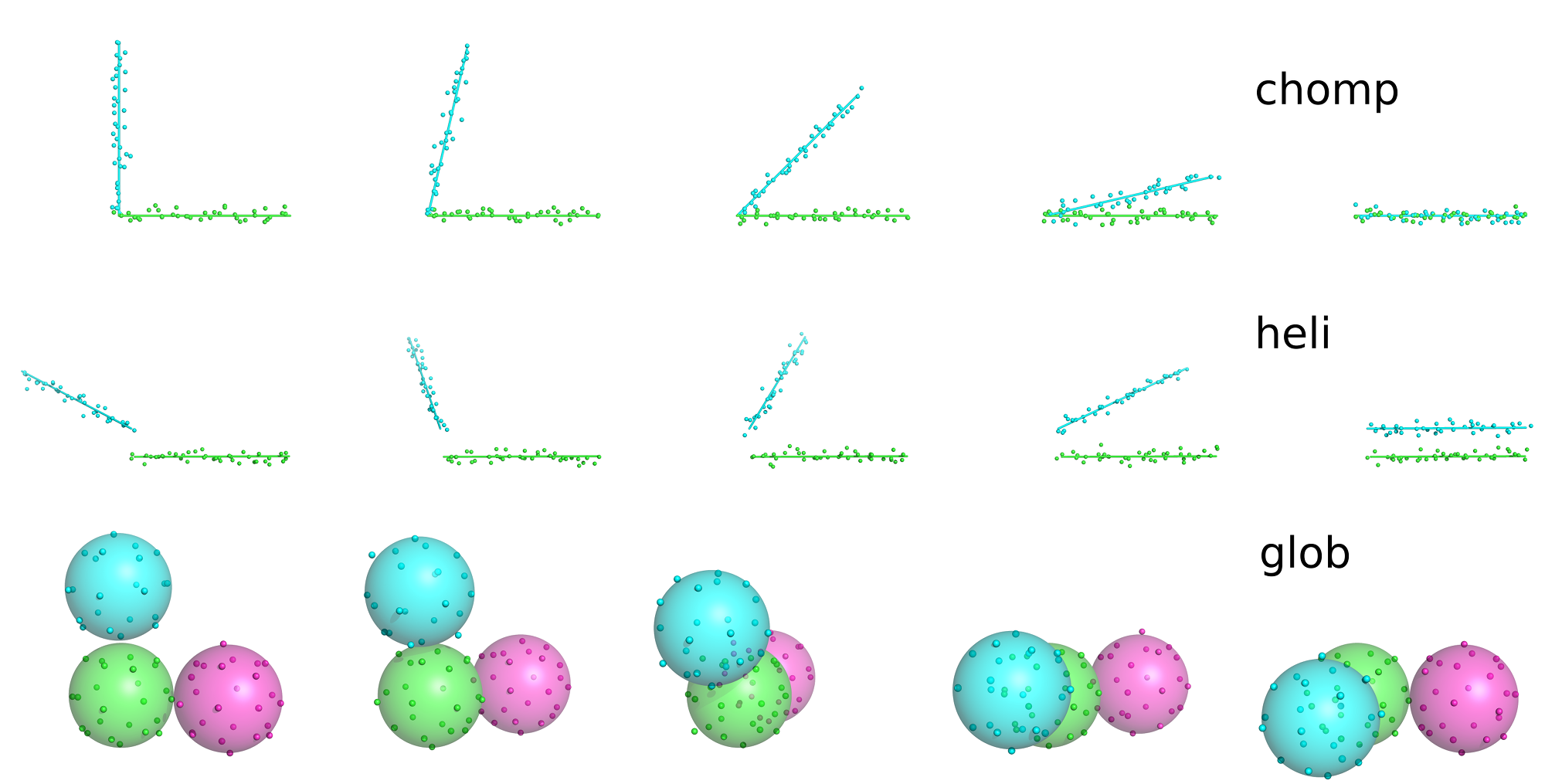}
\caption{Sketch of the three synthetic motions used for testing.
Lines, spheres, and colors are used to
guide the eye, but the classification engine sees
only the unlabeled points.
The time axis proceeds left to right.
For the `glob' system, the sphere on the
right moves behind, and then back to the right again
while the top sphere moves slowly downward.}\label{f:synth}
\end{figure}

Figure~\ref{f:synth} shows images of these three test trajectories.
A complete description of the coordinate generation methods
is present in the Supplementary Information.

  All three systems were processed into binary feature data by
calculating pairwise distances between all points.  Pairs of points
within $2$ distance units were translated to $1$ (representing contact).
When forming the feature vectors, $x$, we removed features, $j$, for
which every sample showed the same result (all contacting or all disconnected).
It is important to note that this removal changes $M$ seen by the algorithm
-- usually decreasing it well below $P (P-1)/2$.

Critically, we repeated these classifications for a range
of material points, $P$.
This tested robustness of the unsupervised classification
problem with respect to the amount of features available.
Adding more points without changing the geometry of the
problem shouldn't change the number of categories detected.

  For each run, five independent MCMC chains were
started from an assignment of all points to a single category.
Each chain ran for 1000 steps.  Samples were collected every 10 steps
-- starting from step 500.
The acceptance probabilities for category split/join
Monte Carlo moves varied around 10-13\%.

\section{ Results}

  We analyzed the results of MC in two different ways.  First,
the categories assigned were tested for grouping in time.
Since the contact lists (on which the categorization
is based) varied slowly over time, we expect categories
to come in `runs.'  Second, we computed histograms
over the number of categories, $K$.  This is a strong
test of the method's sensitivity to the number of material points, $P$.

\begin{table}
\begin{tabular}{l|ll|ll|ll|}
$P$ & 
\multicolumn{2}{c|}{chomp} & 
\multicolumn{2}{c|}{heli} &
\multicolumn{2}{c|}{glob} \\ \hline
18 & 89.2 & 98.0 & 89.9 & 98.9 & 91.1 & 96.7 \\
42 & 94.7 & 98.9 & 93.7 & 99.0 & 93.5 & 97.3 \\
78 & 96.3 & 99.3 & 95.3 & 99.4 & 95.5 & 98.0 \\
114 & 96.8 & 99.4 & 96.5 & 99.5 & 96.4 & 98.4 \\
150 & 97.2 & 99.4 & 96.7 & 99.4 & 97.0 & 98.5 \\
222 & 97.8 & 99.5 & 97.0 & 99.7 & 98.2 & 99.1 \\
258 & 98.4 & 99.7 & 97.7 & 99.5 & 98.6 & 99.4 \\
330 & 98.4 & 99.6 & 97.8 & 99.7 & 98.7 & 99.4 \\
438 & 98.5 & 99.7 & 98.1 & 99.5 & 98.8 & 99.4 \\
\hline
\end{tabular}
\caption{ Categorization accuracy for each system type and number of points, $P$.
The left column for each system is the percent of category assignments
identical to their previous time (not including $K-1$ required).
The right column for each is the percent of category assignments that do not
switch between runs.}\label{t:acc}
\end{table}

\begin{table}
\begin{tabular}{l|ll|ll|ll|}
$N$ & 
\multicolumn{2}{c|}{chomp} & 
\multicolumn{2}{c|}{heli} &
\multicolumn{2}{c|}{glob} \\ \hline
125 & 99.3 & 99.2 & 99.8 & 99.2 & 99.5 & 99.3 \\
250 & 99.4 & 99.6 & 99.3 & 99.6 & 99.2 & 99.4 \\
500 & 98.8 & 99.6 & 97.7 & 99.7 & 98.6 & 99.3 \\
1000 & 97.7 & 99.6 & 96.9 & 99.6 & 98.2 & 99.2 \\
2000 & 96.9 & 99.4 & 96.5 & 99.6 & 98.0 & 99.1 \\
\hline
\end{tabular}
\caption{ Categorization accuracy for each test system
as the sample number, $N$, varies.
The number of points remains fixed at $P=222$.
Column labels are as in Table~\ref{t:acc}.}\label{t:acc2}
\end{table}

  As expected, we found a high degree of correlation between
categories and time for every case.
Similar time-points were grouped into similar
categories.  To quantify these results, we counted transitions between category
indices in time-order.  For a perfect categorization, the number
of transitions should equal the number of categories minus one.
We computed the categorization accuracy in two ways.
For each system, the left columns of
tables~\ref{t:acc} and~\ref{t:acc2}
are 100 minus the percent of mis-categorized frames.
For lossy categorization at time-boundaries, we expect
oscillation between two values.  We quantified this by forming
a transition matrix between categories, and removing
transitions along the `most likely path'.  Excluding this boundary
oscillation we found nearly 100\% accuracy for the categorizations.
Those are shown in the right columns of tables~\ref{t:acc}
and~\ref{t:acc2}.

\begin{figure*}
\centering
 \subfloat[Fixed $N=1000$]{
   \includegraphics[width=0.45\textwidth]{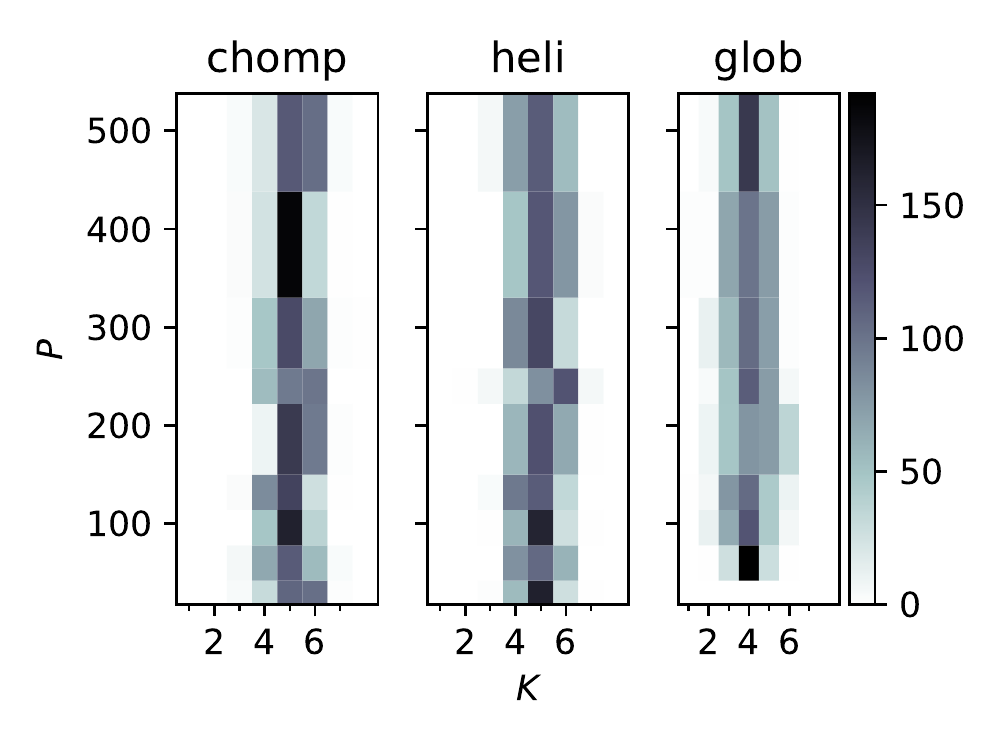}\label{f:PK}
 }
 \subfloat[Fixed $P=222$]{
   \includegraphics[width=0.45\textwidth]{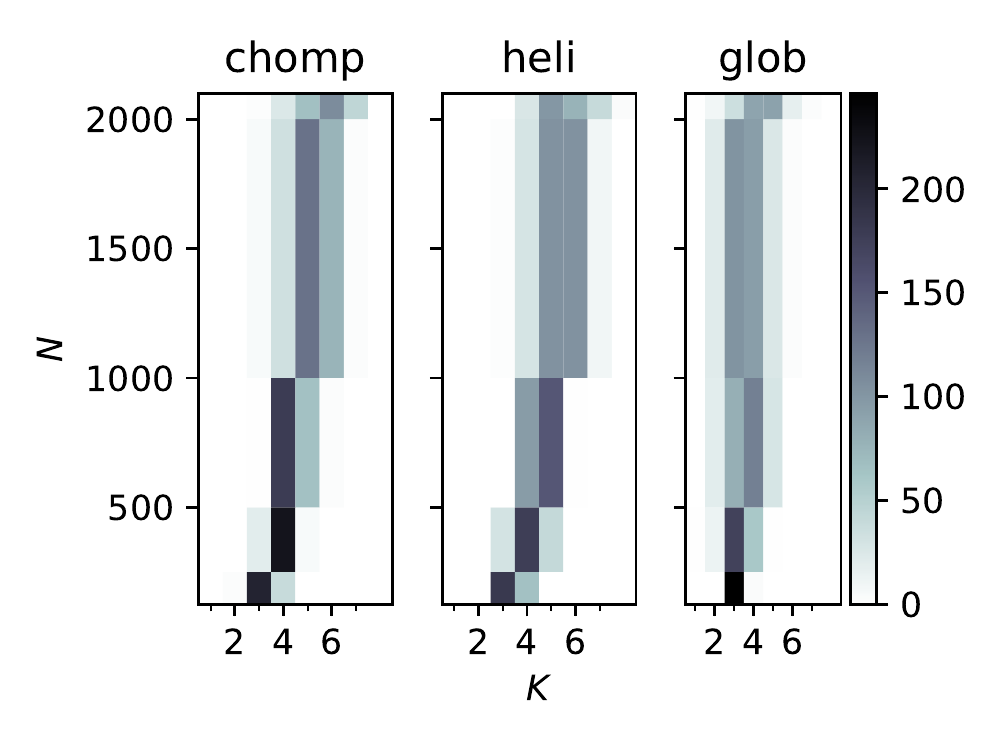}\label{f:NK}
 }
\caption{Counts (out of 250 samples each) of the number of categories, $K$,
as $P$ or $N$ are varied.  The shape of the distribution
function remains essentially constant
as $P$ varies, but tends to spread toward higher $K$ with more samples.}\label{f:ncat}
\end{figure*}

  Integrating the posterior probability (Eq.~\ref{e:like} times Eq.~\ref{e:prior})
over $\pi$ leads to factors like $\Gamma(N+\alpha K)^{-1}$,
which seem prohibitively costly as $N$ increases.  We therefore wanted
to check that the number of categories doesn't decrease as features or
samples are added.  Fig.~\ref{f:ncat} shows the sampled probability
distributions over $K$, the number of categories, for increasing $P$~(\ref{f:PK})
and $N$~(\ref{f:NK}).
Interestingly, for every system, about five conformational states were deduced
at $N=1000$.
As $N$ increased, however, more categories were deduced and the distribution
spread to higher numbers.
This is probably reasonable, since more values of the `time' coordinate
generated a more fine-grained motion.

\subsection{ Adenylate kinase open/closed transition}

  Finally, we tested the classification method against a protein
with a well-characterized conformational transition.  Adenylate kinase (ADK)
converts ATP, ADP, and AMP by closing around substrate molecules.\cite{cmull92}
The transition from closed to open was simulated in Ref.~\cite{jvant20}
using steered molecular dynamics on a reaction coordinate
interpolating between the electron density
maps of PDB IDs 1AKE (Ref.~\cite{cmull92}, closed) and 4AKE (Ref.~\cite{cmull96}, open).
The simulation data we used did not contain ligands, but did
contain water and ions.  Our analysis only made use of the
alpha carbon (C$_\alpha$) positions.

  Features were calculated for each of $N=3900$ equally spaced
frames during steered dynamics by testing whether C$_\alpha$ to
C$_\alpha$ distances were less than 5~\AA{}.  These structures
contain $P=214$ C$_\alpha$-s.  Sampling was
carried out as described in Sec.~\ref{s:test}, but 8 independent
MC chains were sampled for 1250 steps (instead of 5 for 1000).
The acceptance probability of split/join moves was 17\%.
During sampling, we saved the parameters, $\theta$,
possessing maximum likelihood values at each $K$.

\begin{figure*}
\centering
 \subfloat[][]{
   \includegraphics[width=0.4\textwidth]{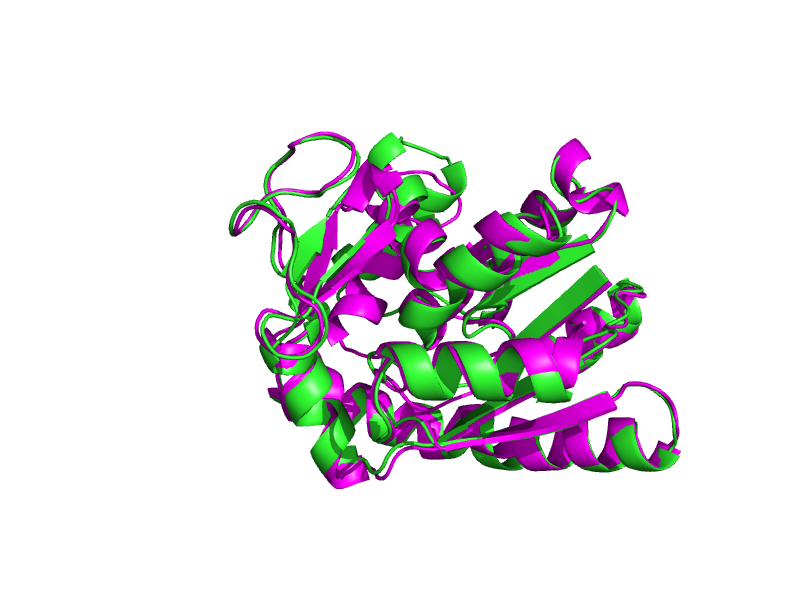}\label{f:closed}
 }
 \subfloat[][]{
   \includegraphics[width=0.4\textwidth]{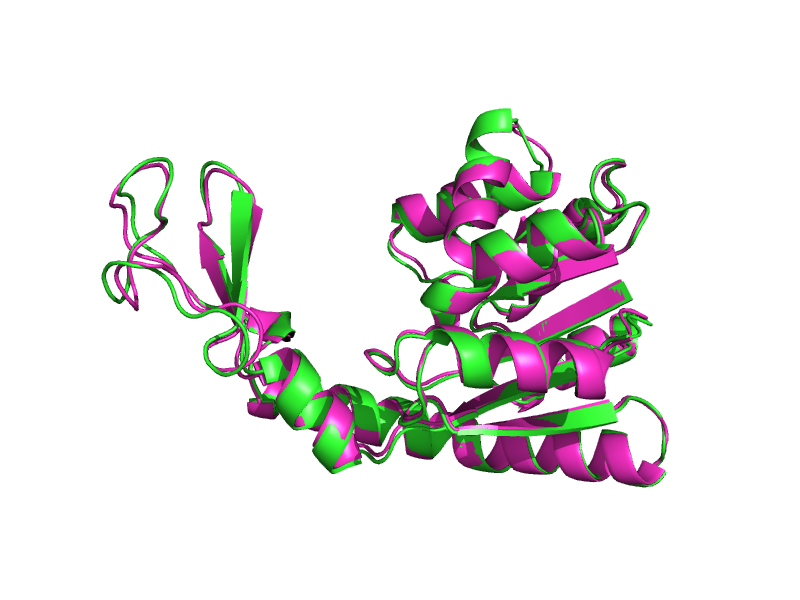}\label{f:open}
 }
\caption{Adenylate kinase (ADK) closed~\protect\subref{f:closed} and open~\protect\subref{f:open} 
configurations.  PDB structures (magenta) are compared to the most similar
molecular dynamics classification results (green)
found among the five representative structures for the $K=5$ classification.
}\label{f:adk}
\end{figure*}

  We then extracted conformational states with the highest
probability for landing in each category as representative
points for that category.  Since the reference open and closed PDB
conformations formed extreme points, our representative
structures approached them more nearly as $K$ increased.
Fig.~\ref{f:adk} shows that the two end-point conformations
ended up very close to the open and closed states from the PDB.

\section{ Conclusion}


  The method developed here is ideally suited for the unsupervised structural classification
problem.  It has been derived from a first-principles Bayesian analysis of
the set of atoms which interact within a structure.  This work solved the
central problem of defining an appropriate prior distribution over parameter
space and implementing an efficient sampling method.

  Tests on sample conformational transitions
identified more categories than na\"{\i}vely expected
because it generated milestones along the motion's time-coordinate.
However, all categorizations were shown to have
excellent accuracy as judged by picking out the correct time-sequence.
Finally, a test on Adenylate kinase verified that these conclusions
easily generalize to protein motions.

  The central results of this work showed that the method
behaves well under large variations in the number of features
and samples.  These observations validate our choice for the prior
distribution, since changes in $P(\theta|I)$ will have large
effects on the distribution over number of categories, $P(K|x I)$.
They also show robustness of the MC sampling method itself,
since relatively high acceptance rates were achieved.

  Many future applications and development of this method
are possible.  Changes in the pair contacts between states
could be analyzed more thoroughly, as in the DynDom method.\cite{rlee03}
Probabilities of assignment to each conformational state can be
used as reaction coordinates.  We are presently applying
the method to classify chemical compound
space using binary MACCS fingerprints\cite{maccs},
and to characterize conformational space of SARS-CoV-2 proteins
simulated using replica exchange molecular dynamics.\cite{covid}


  Now that the principle has been demonstrated, more informative
classification schemes can be devised.  Adding information
like hydrogen-bonding, salt bridge formation, and secondary
structure annotation will allow the Bayesian framework
able to recognize categories more like a biochemist would.
The method can also be focused on active domains
and binding sites by adding more points and shorter distance
cutoffs for key residues.  The insensitivity to $M$
shown in this work provides a high degree of confidence that any amount of
additional data will improve the overall categorization without spoiling
the classification already achieved using coarser-level data. 


\begin{thebibliography}{10}

\bibitem{jguo16}
Jingjing Guo and Huan-Xiang Zhou.
\newblock Protein allostery and conformational dynamics.
\newblock {\em Chemical Reviews}, 116(11):6503--6515, February 2016.

\bibitem{cmull96}
CW~M\"{u}ller, GJ~Schlauderer, J~Reinstein, and GE~Schulz.
\newblock Adenylate kinase motions during catalysis: an energetic counterweight
  balancing substrate binding.
\newblock {\em Structure}, 4(2):147--156, February 1996.

\bibitem{camar12}
C.~Amaral, V.~Carnevale, M.~L. Klein, and W.~Treptow.
\newblock Exploring conformational states of the bacterial voltage-gated sodium
  channel {NavAb} via molecular dynamics simulations.
\newblock {\em Proc. Nat. Acad. Sci. USA}, 109(52):21336--21341, November 2012.

\bibitem{ykim13}
Yujin~E. Kim, Mark~S. Hipp, Andreas Bracher, Manajit Hayer-Hartl, and F.~Ulrich
  Hartl.
\newblock Molecular chaperone functions in protein folding and proteostasis.
\newblock {\em Annu. Rev. Biochem.}, 82(1):323--355, June 2013.

\bibitem{jvant20}
John~W. Vant, Daipayan Sarkar, Giacomo Fiorin, Robert Skeel, Josh~V. Vermaas,
  and Abhishek Singharoy.
\newblock Data-guided multi-map variables for ensemble refinement of molecular
  movies.
\newblock 2020.
\newblock \href{https://www.biorxiv.org/content/10.1101/2020.07.23.217794v1.abstract}{submitted}.

\bibitem{rlee03}
Richard~A. Lee, Moe Razaz, and Steven Hayward.
\newblock The {DynDom} database of protein domain motions.
\newblock {\em Bioinformatics}, 19(10):1290--1291, 2003.

\bibitem{cgird15}
Christopher Girdlestone and Steven Hayward.
\newblock The {DynDom3D} webserver for the analysis of domain movements in
  multimeric proteins.
\newblock {\em J. Comput. Biol.}, pages 21--26, 2016.

\bibitem{hstam10}
Hern\'{a}n Stamati, Cecilia Clementi, and Lydia~E. Kavraki.
\newblock Application of nonlinear dimensionality reduction to characterize the
  conformational landscape of small peptides.
\newblock {\em Proteins: Structure, Function, and Bioinformatics},
  78(2):223--235, 2010.

\bibitem{arama11}
Arvind Ramanathan, Andrej~J. Savol, Christopher~J. Langmead, Pratul~K. Agarwal,
  and Chakra~S. Chennubhotla.
\newblock Discovering conformational sub-states relevant to protein function.
\newblock {\em {PLoS} {ONE}}, 6(1):e15827, January 2011.

\bibitem{aferg11}
Andrew~L. Ferguson, Athanassios~Z. Panagiotopoulos, Ioannis G.Kevrekidis, and
  Pablo G.Debenedetti.
\newblock Nonlinear dimensionality reduction in molecular simulation: The
  diffusion map approach.
\newblock {\em Chemical Physics Letters}, 509(1--3):1--11, 2011.

\bibitem{dmoji13}
Mojie Duan, Jue Fan, Minghai Li, Li~Han, and Shuanghong Huo.
\newblock Evaluation of dimensionality-reduction methods from peptide
  folding{\textendash}unfolding simulations.
\newblock {\em J. Chem. Theory. Comput.}, 9(5):2490--2497, April 2013.

\bibitem{bspiw20}
Vojt\v{e}ch Spiwok and Pavel K\v{r}\'{\i}\v{z}.
\newblock Time-lagged t-distributed stochastic neighbor embedding ({t-SNE}) of
  molecular simulation trajectories.
\newblock {\em Frontiers in Molecular Biosciences}, 7:132, 2020.

\bibitem{cli16}
Cheng Li, Bingyu Wang, Virgil Pavlu, and Javed Aslam.
\newblock Conditional {Bernoulli} mixtures for multi-label classification.
\newblock In {\em Proc. 33$^\mathrm{rd}$ International Conf. on Machine
  Learning}, volume~48, 2016.

\bibitem{valab06}
Vicent Alabau, Jes\'{u}s Andr\'{e}s, Francisco Casacuberta, Jorge Civera,
  Jos\'{e} Garc\'{\i}a-Hern\'{a}ndez, Adri\`{a} Gim\'{e}nez, Alfons Juan,
  Alberto Sanchis, and Enrique Vidal.
\newblock The naive {Bayes} model, generalisations and applications.
\newblock In Filiberto Pla, Petia Radeva, and Vitri\`{a}, editors, {\em Pattern
  Recognition: Progress, Directions and Applications}, 2006.

\bibitem{jnovo04}
Jana Novovi\v{c}ov\'{a} and Anton\'{\i}n Mal\'{\i}k.
\newblock Text document classification based on mixture models.
\newblock {\em Kybernetika}, 40(3):293--304, 2004.

\bibitem{dkaji10}
Daisuke Kaji, Kazuho Watanabe, and Sumio Watanabe.
\newblock Phase transition of variational bayes learning in {Bernoulli}
  mixture.
\newblock {\em Australian Journal of Intelligent Information Processing
  Systems}, 11(4):35--40, 2010.

\bibitem{kyama13}
Keisuke Yamazaki and Daisuke Kaji.
\newblock Comparing two {Bayes} methods based on the free energy functions in
  {Bernoulli} mixtures.
\newblock {\em Neural Networks}, 44:36--43, 2013.

\bibitem{ejayn03}
E.~T. Jaynes.
\newblock {\em Probability Theory, The Logic of Science}.
\newblock Cambridge University Press, Cambridge, 2003.

\bibitem{cmull92}
Christoph~W. M\"{u}ller and Georg~E. Schulz.
\newblock Structure of the complex between adenylate kinase from escherichia
  coli and the inhibitor ap5a refined at 1.9~\aa{} resolution.
\newblock {\em J. Mol. Biol.}, 224(1):159--177, March 1992.

\bibitem{maccs}
Joseph~L. Durant, Burton~A. Leland, Douglas~R. Henry, and James~G. Nourse.
\newblock Reoptimization of {MDL} keys for use in drug discovery.
\newblock {\em J. Chem. Inf. Comput. Sci.}, 42(6):1273--1280, 2002.

\bibitem{covid}
A.~Acharya, R.~Agarwal, M.~Baker, J.~Baudry, D.~Bhowmik, S.~Boehm, K.~Byler,
  L.~Coates, S.Y. Chen, C.J. Cooper, O.~Demerdash, I.~Daidone, J.D. Eblen,
  S.~Ellingson, S.~Forli, J.~Glaser, J.~C. Gumbart, J.~Gunnels, O.~Hernandez,
  S.~Irle, J.~Larkin, T.J. Lawrence, S.~LeGrand, S.-H. Liu, J.C. Mitchell,
  G.~Park, J.M. Parks, A.~Pavlova, L.~Petridis, D.~Poole, L.~Pouchard,
  A.~Ramanathan, D.~Rogers, D.~Santos-Martins, A.~Scheinberg, A.~Sedova,
  S.~Shen, J.C. Smith, M.D. Smith, C.~Soto, A.~Tsaris, M.~Thavappiragasam, A.F.
  Tillack, J.V. Vermaas, V.Q. Vuong, J.~Yin, S.~Yoo, M.~Zahran, and
  L.~Zanetti-Polzi.
\newblock Supercomputer-based ensemble docking drug discovery pipeline with
  application to {Covid-19}.
\newblock 2020.
\newblock \href{https://chemrxiv.org/articles/preprint/Supercomputer-Based\_Ensemble\_Docking\_Drug\_Discovery\_Pipeline\_with\_Application\_to\_Covid-19/12725465}{submitted}.

\end{thebibliography}

\section*{ Acknowledgment}

We thank Josh Vermaas for providing simulated trajectories of ADK
and helpful comments on the manuscript.
Research sponsored by the Laboratory Directed Research and Development Program
of Oak Ridge National Laboratory (ORNL).
This research used resources of the Oak Ridge Leadership Computing Facility at ORNL.
ORNL is managed by UT-Battelle, LLC, for the US
Department of Energy under contract DE-AC05-00OR22725.

\onecolumn
\section*{ Supplementary Information}

The supplementary information accompanying this article contains a small
proof of the scaling of the log-posterior probability distribution with $M$,
showing its connection to information entropy.  It also contains
an explicit formula for the generation and acceptance probabilities
of split and join MC moves.  Finally, a full description of the
synthetic geometries is provided.
The source code for this classification method, along with the code
to reproduce the test calculations in this paper,
is publicly available from
\href{https://github.com/frobnitzem/classifier}{https://github.com/frobnitzem/classifier}.

\section{ Supplementary Information}

\subsection{ Behavior of the posterior PDF with increasing $M$}

  We argue here that the implicit uniform prior for all $\mu_{kj}$
as well as the choice $\alpha = M+1$ is most appropriate
for this problem.  This is the conclusion we arrive at by
requiring that the likelihood for dividing a category into two parts should be
insensitive to $M$.

  First, multiply $P(xz|\theta)$ and $P(\theta|U)$ from the main text to provide
 $P(xz\theta | U)$,
\begin{equation}
\frac{\Gamma(K \alpha)}{\Gamma(\alpha)^K}
\prod_{k=1}^K
\left(\frac{\Gamma(2\beta)}{\Gamma(\beta)^2}\right)^M
\pi_k^{N_k+\alpha-1} \prod_{j=1}^M
\mu_{kj}^{N_{kj}+\beta-1} (1-\mu_{kj})^{N_k-N_{kj}+\beta-1}
.\label{e:post}
\end{equation}
This is generalized a bit from the main text so that a Beta prior for $\mu$ (with parameter $\beta$) is explicitly shown.

  Observe that (without the uniqueness constraint)
the posterior distribution over categories is proportional to,
\begin{equation}
P(xz|U) \propto f(N,K)
\prod_{k} \Gamma(N_k + \alpha) \prod_j \frac{\Gamma(N_{kj}+\beta)\Gamma(N_k-N_{kj}+\beta)}{\Gamma(N_k+2\beta)}
,
\end{equation}
with $f(N,K) = \Gamma(\alpha K) / \Gamma(\alpha)^K \Gamma(N+\alpha K)$
(for $\beta=1$).
The probability per category on the far-right can be likened to an information entropy,
\begin{equation}
S_\text{inf}(kj) \equiv \ln \binom{N_k}{N_{kj}}
,
\end{equation}
but is offset because of the limit,
\begin{equation}
\lim_{n\to\infty} \frac{\Gamma(n+a)}{\Gamma(n) n^a} = 1
.\label{e:lim}
\end{equation}

Inserting this limit on the right, we find
\begin{equation}
\frac{\Gamma(N_{kj}+\beta)\Gamma(N_k-N_{kj}+\beta)}{\Gamma(N_k+2\beta)}
	\approx e^{-S_\text{inf}(kj)}
	\frac{N_{kj}^{\beta-1}(N_k-N_{kj})^{\beta-1}}{N_k^{2\beta-1}}
.
\end{equation}
This provides the motivation for choosing $\beta = 1$.
This choice for $\beta$ makes the categorization closer to an
information entropy, depending only on the ratio $N_{kj}/N_k$.

The categorization probability
is now asymptotically like an information entropy, except for the extra
factor of $1/(N_k+1)$.  This factor is present for every feature, and creates an undesired
dependence on $M$.
We can cancel that factor by appealing to Eq.~\ref{e:lim}.
This time using $n = N_k+1$ and $a = \alpha-1$ to approximate
$\Gamma(N_k+\alpha)$,
\begin{equation}
P(xz|U) \approx f(N,K)
\prod_{k} \frac{N_k! (N_k+1)^{\alpha-1}}{(N_k+1)^M} \prod_j e^{-S_\text{inf}(kj)}
. \label{e:pri2}
\end{equation}

  Eq.~\ref{e:pri2} provides the motivation for choosing $\alpha = M+1$.
With this choice, the $(N_k+1)^M$ term cancels.  It leaves behind
a relative information entropy for the $N_k$.  Essentially,
\begin{equation}
P(xz|U) \approx f(N,K) N!
\exp{ \left( S_\text{inf}(\{N_k\}) - \sum_{kj} S_\text{inf}(kj) \right) }
. \label{e:pri3}
\end{equation}

  At this point, different choices for $f(N,K)$ lead to differences in
the probability distribution over $K$, the number of categories.
Rather than directly influencing these, we have chosen not
to create additional assumptions about $f(N,K)$ beyond the above
choices for $\alpha$ and $\beta$.

\subsection{ Sampling Method}

  Since $\pi$ can be trivially integrated out of the posterior distribution
(Eq.~\ref{e:post}), we re-generate $\pi$ on every split/join move,
and consider only the integrated probability $P(z\mu|xI)$ as our sampling target.
Each time a split/join move is generated,  a re-categorization of
samples in category $k$ ($\{x_k\}$) into $\{x_L\}$ and $\{x_R\}$.
is immediately followed by generation of $\mu_L$ and $\mu_R$ (or $\mu_k$ for a join)
from their respective Beta distributions (with samples $N_{Lj}$ and $N_L-N_{Lk}$, etc.).

  We set up the acceptance probability for these split / join moves
to satisfy the detailed balance condition,
\begin{equation}
\frac{P_\text{acc}(z'\mu' | z\mu x)}{P_\text{acc}(z\mu | z'\mu' x)}
  = \frac{P_\text{gen}(z\mu |z' \mu' x)}{P_\text{gen}(z'\mu'| z \mu x)}
    \frac{P(z'\mu'|xI)}{P(z\mu |x I)}
    .\label{e:acc}
\end{equation}
Without loss of generality, assume $z'$ contains $K+1$ categories and $z$
contains $K$.
We already have explicit formulas for $P(z\theta |x I)$ (modulo $P(x|I)$),
so the only thing needed to calculate $P_\text{acc}$
is the move generation probabilities.

  Given category $k$ and feature $j$ was chosen to define the split (see main text for
a description of this process), the recategorization probability is,
\begin{equation}
P_\text{gen}(z'|z kj) = \frac{1}{2} \left(
\begin{split}
 &\eta^{N_{Lj}+N_R-N_{Rj}} (1-\eta)^{N_{Rj}+N_L-N_{Lj}} \\
 &\quad + (1-\eta)^{N_{Lj}+N_R-N_{Rj}} \eta^{N_{Rj}+N_L-N_{Lj}}
\end{split}
\right) / (1-p_L-p_R)
.\label{e:Psplit}
\end{equation}
The factor of $\tfrac{1}{2}$ is necessary because the labels $L$ and $R$ are interchangeable.
$p_L$ ($p_R$) is the expression in the numerator when all $z'$ move from $k$ to $L$ (R).

  Because any $j$ could have been chosen, the overall probability of ending up
with categories $L$ and $R$ is,
\begin{equation}
P_\text{gen}(z'|z) = \frac{\xi(K)}{\sum_{nl} Q_{nl}} \sum_j Q_{kj} P_\text{gen}(z'|z kj)
.
\end{equation}
To concentrate splitting on productive cases, we choose $Q_{kj}$
to be one for all $k,j$ unless $N_{kj} = 0$ or $N_{kj} = N_k$.
In those cases $Q_{kj}$ is zero so that we don't split using a non-informative feature.

  Generating $\mu_L$ and $\mu_R$ adds the factor,
\begin{equation}
P_\text{gen}(z'\mu'|zx) = P_\text{gen}(z'|z)
  P(\{x_L\}|\mu_L) (N_L+1)^M P(\{x_R\}|\mu_L) (N_R+1)^M
  e^{\sum_{j} S_\text{inf}(Lj)+S_\text{inf}(Rj)}
  .
\end{equation}
Here, we have made the intuitive definition,
\begin{equation}
P(\{x_k\}|\mu_k) \equiv \prod_j \mu_{kj}^{N_{kj}} (1-\mu_{kj})^{N_k-N_{kj}}
.
\end{equation}
Most of this term will cancel $P(z\mu|xI)$ in the final expression for the acceptance probability.

  The recombination rule is attempted with frequency $1-\xi(K+1)$.
A set $u$ is chosen at random with probability $1/(K+1)$ and
recombined with set $v \ne u$ chosen with probability
$2/K(K+1)$.
The resulting generation probability is,
\begin{equation}
P_\text{gen}(z|z'x) = \frac{(1-\xi(K+1)) 2}{K (K+1)}
.
\end{equation}

  Generating $\mu_k$ adds the factor,
\begin{equation}
P_\text{gen}(z\mu|z'x) = P_\text{gen}(z|z'x)
  P(\{x_k\}|\mu_k) (N_k+1)^M
  e^{\sum_{j} S_\text{inf}(kj)}
  .
\end{equation}

  As a final aid to computing the acceptance probability ratio (Eq.~\ref{e:acc}),
we provide a simplified expression for the likelihood ratio,
\begin{align}
\frac{P(z'\mu' | x I)}{P(z\mu | x I)} &=
\frac{ (1-B_{LR}) \prod_{k' \ne L,R} \left[ (1-B_{Lk'}) (1-B_{Rk'}) \right] }
       {\prod_{k' \ne k} (1-B_{kk'})}
\frac{\Gamma(\alpha(K+1))}{\Gamma(\alpha K)\Gamma(\alpha)}
\frac{\Gamma(N+\alpha K)}{\Gamma(N+\alpha (K+1))} \notag \\
 &\quad \frac{\Gamma(N_L+\alpha)\Gamma(N_R+\alpha)}{\Gamma(N_k+\alpha)}
\frac{P(\{x_L\} | \mu_L) P(\{x_R\} | \mu_R) }{P(\{x_k\} | \mu_k)}
\end{align}

\subsection{ Coordinate Trajectory Generation for Synthetic Test Systems}

The `chomp' system was 2D, with two line segments, both of length 4 units.
The bottom segment ($P/2$ uniformly spaced particles) stretched along the $x$-axis
and remained stationary.  The top segment ($P/2-1$ uniformly spaced particles,
since the origin is excluded),
moved at angle $\theta(t) = \frac{\pi}{4} \cos(t)$, $t \in [0,\pi]$.
This motion mimicks a harmonic oscillator -- which spends a majority
of its time at the two limits of its motion.

The `heli' system was 3D. It also had two line segments
with $P/2$ equally spaced particles each.  The bottom segment remained as before,
while the top segment was replaced by $P/2$ particles displaced from
the bottom one by 1 distance unit along $z$.
The top segment was rotated in the $xy$ plane
with $\theta(t) = t$, $t \in [0,\pi]$.

The `glob' system was 3D, consisting of $3$ spheres, each represented
by $P/3$ points at radius 2.  The point locations were taken from Lebedev quadrature
rules, which possess octahedral symmetry.
Sphere 1 remained fixed at the origin.
Spheres 2 and 3 started in locations displaced along $z$ or $x$, respectively,
by 2.1 distance units.
As $t$ traversed $[0,\pi]$, sphere 2 was rotated about the origin
around the $x$-axis by angle
$\theta_2(t) = \frac{\pi}{4} (1 - \cos(t))$, while
sphere 3 was rotated about the origin around the $z$-axis by angle
$\theta_3(t) = \frac{\pi}{4} (1 - \cos(2t))$.
This motion creates something like three conformational states.
Since sphere 3 moves twice as fast as sphere 2, sphere 2 can be at
$+z$ or $-y$ while sphere 3 is at $+x$.  A third conformation is at the
center where sphere 3 is at $+y$.

\end{document}